\documentclass[iop,apj,tighten,twocolappendix,numberedappendix]{emulateapj}
\usepackage{apjfonts}

\usepackage{graphicx}
\usepackage{amsmath}
\usepackage{amssymb}
\usepackage{color}
\usepackage{mathtools}
\usepackage{url}

\usepackage[breaklinks,colorlinks,citecolor=blue,linkcolor=red]{hyperref} 
\usepackage[all]{hypcap}

\newcommand\msun{\, \rm M_\odot}
\newcommand\rsun{\, \rm R_\odot}
\newcommand\kms{\, \rm km\,s^{-1}}

\newcommand\vkick{{v_{\rm kick}}}

\newcommand\be{\begin{equation}}
\newcommand\ee{\end{equation}}

\begin{document}

\title{Merging Black Holes in the Low-mass and High-mass Gaps from 2+2 Quadruple Systems}

\author{Giacomo Fragione\altaffilmark{1,2}, Abraham Loeb \altaffilmark{3}, \& Frederic A.\ Rasio\altaffilmark{1,2}}
 \affil{$^1$Department of Physics \& Astronomy, Northwestern University, Evanston, IL 60208, USA} 
  \affil{$^2$Center for Interdisciplinary Exploration \& Research in Astrophysics (CIERA)}
  \affil{$^3$Astronomy Department, Harvard University, 60 Garden St., Cambridge, MA 02138, USA}

\begin{abstract}
The origin of the black hole (BH) binary mergers observed by LIGO-Virgo is still uncertain, as are the boundaries of the stellar BH mass function. Stellar evolution models predict a dearth of BHs both at masses $\gtrsim 50\msun$ and $\lesssim 5\msun$, thus leaving low- and high-mass gaps in the BH mass function. A natural way to form BHs of these masses is through mergers of neutron stars (NSs; for the low-mass gap) or lower-mass BHs (for the high-mass gap); the low- or high-mass-gap BH produced as a merger product can then be detected by LIGO-Virgo if it merges again with a new companion. We show that the evolution of a 2+2 quadruple system can naturally lead to BH mergers with component masses in the low- or high-mass gaps. In our scenario, the BH in the mass gap originates from the merger of two NSs, or two BHs, in one of the two binaries and the merger product is imparted a recoil kick (from anisotropic gravitational wave emission), which triggers its interaction with the other binary component of the quadruple system. The outcome of this three-body interaction is usually a new eccentric compact binary containing the BH in the mass gap, which can then merge again. The merger rate is $\sim 10^{-7} - 10^{-2}$ Gpc$^{-3}$ yr$^{-1}$ and $\sim 10^{-3} - 10^{-2}$ Gpc$^{-3}$ yr$^{-1}$ for BHs in the low-mass and high-mass gap, respectively. As the sensitivity of gravitational wave detectors improves, tighter constraints will soon be placed on the stellar BH mass function.
\end{abstract}

\keywords{stars: kinematics and dynamics – stars: neutron – stars: black holes – Galaxy: kinematics and dynamics}

\section{Introduction}
\label{sect:intro}

The existence of stellar-mass black holes (BHs) has been proven beyond any reasonable doubt by LIGO-Virgo observations of $10$ BH--BH binary mergers \citep{ligo2018}. However, the likely formation mechanisms for these mergers are still highly uncertain. Several candidates could potentially account for most of the observed events, including mergers from isolated binary star evolution \citep{bel16b,demi2016,brei2019,spera2019}, dynamical formation in dense star clusters \citep{askar17,baner18,fragk2018,rod18,sams18,ham2019,krem2019}, mergers in triple and quadruple systems induced through the Kozai-Lidov (KL) mechanism \citep{antoper12,anm14,alk2018,ll18,fragg2019,flp2019,fragk2019,liu2019}, mergers of compact binaries in galactic nuclei \citep{bart17,sto17,rasskoc2019,mck2020}, and mergers of primordial black holes \citep{sasaki2016}.

Also highly uncertain are the exact boundaries of the BH mass function \citep{perna2019}. Current stellar evolution models predict a dearth of BHs both with masses $\gtrsim 50\msun$ and $\lesssim 5\msun$, based on the details of the progenitor collapse. The high-mass gap results from pulsational pair-instabilities affecting the massive progenitors. These can lead to large amounts of mass being ejected whenever the pre-explosion stellar core is approximately in the range $45\msun - 65\msun$, leaving a BH remnant with a maximum mass around $40\msun - 50\msun$ \citep{heger2003,woosley2017}. The lower boundary of the high-mass gap is estimated to be around $70\msun$ for Pop III stars \citep{woosley2017}, $80\msun$ for intermediate-metallicity stars \citep{limongi2018}, and $70\msun$ for high-metallicity stars \citep{bel2020}; its upper boundary is thought to be around $125\msun$ \citep{renzo2020}. On the other hand, the low-mass gap is related to the explosion mechanism in a core-collapse supernova \citep[SN; see][]{belc2012,fryer2012}. At even lower masses, $\lesssim 3\msun$, neutron stars (NSs) are thought to populate the mass spectrum of compact remnants from stellar collapse. The most massive NS observed to date is about $2.1\msun$ \citep{crom2020}.

A natural way to form BHs both in the low- and high-mass gap is through mergers of NSs and lower-mass BHs, respectively. To detect such BHs through gravitational wave (GW) emission, the merger remnant has to acquire a new companion to merge with. This immediately excludes isolated binaries as a progenitor, thus favoring a dynamical channel. A fundamental limit for repeated mergers in star clusters comes from the recoil kick imparted to merger remnants through anisotropic GW emission \citep{lou10,lou11}. Depending on the mass ratio and the spins of the merging objects, the recoil kick can often exceed the local escape speed, leading to ejection from the system and thus preventing a second merger in the mass gap \citep{gerosa2019}. For NS--NS mergers that could produce BHs in the low-mass gap, the GW recoil kicks are typically less strong since the encounter takes place at a larger gravitational radius than BH--BH mergers, but hydrodynamic effects could become important instead \citep{shibata2005,rezz2010}. For BHs, a number of studies have shown that massive globular clusters \citep{rodri2019}, nuclear clusters \citep{antonini2019}, and AGN disks \citep{mck2020} are the only environments where second-generation mergers can take place, owing to their high escape speed. For NSs, detailed calculations show that NS--NS mergers are so rare in globular clusters that the retention and second merger of a resulting low-mass-gap BH is extremely unlikely \citep{ye2020}.

\begin{figure*} 
\centering
\includegraphics[scale=0.245]{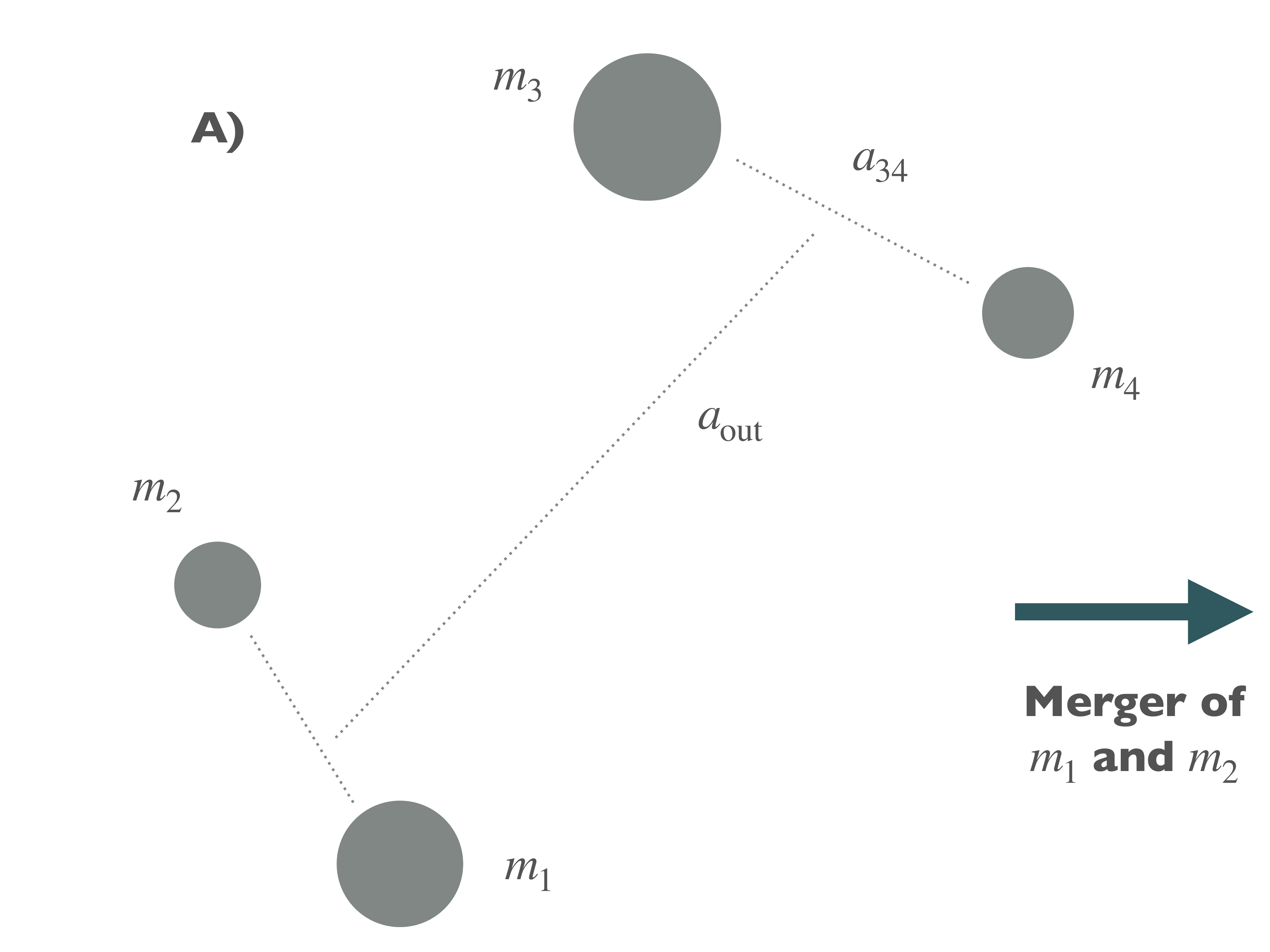}
\includegraphics[scale=0.245]{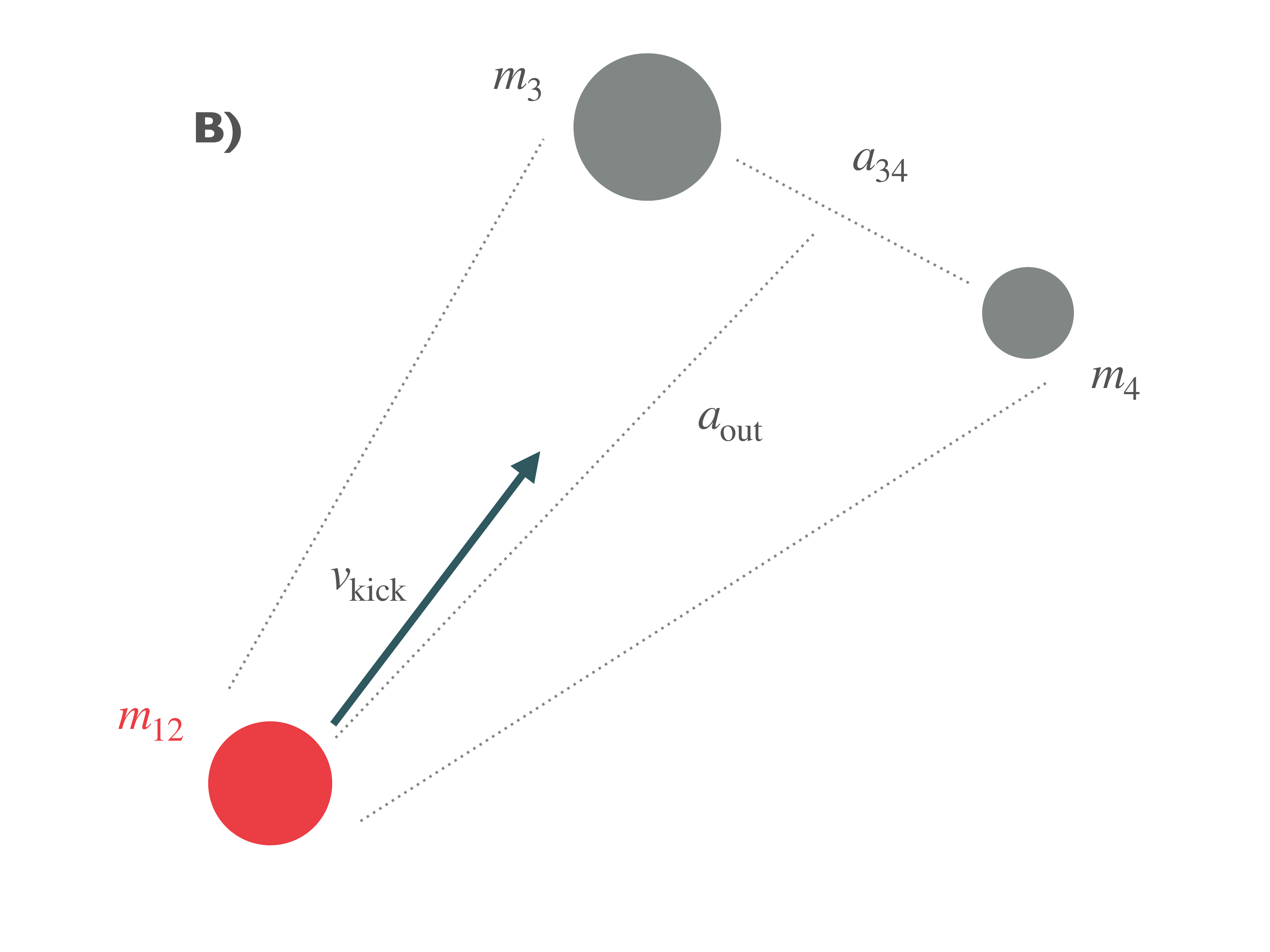}
\caption{Interacting quadruples. Two NSs or BHs in the 2+2 quadruple ($m_1$ and $m_2$) merge producing a BH in either the low- or high-mass gap. The merger remnant is imparted a recoil kick velocity $v_{\rm kick}$, which triggers its interaction with the second binary ($m_3$ and $m_4$). The outcome of the encounter will eventually be a new binary containing the BH in the mass gap, which then can merge again with either $m_3$ or $m_4$.}
\label{fig:interq}
\end{figure*}

Bound stellar multiples are common in the Universe. Observations have shown that the fraction of massive stars, progenitors of NSs and BHs, that have at least one or two stellar companions
is $\sim 50$\% and $\sim 15$\%, respectively \citep{sa2013AA,tok14a,tok14b,duns2015,moe2017,sana2017,jim2019}. Quadruple systems are also observed and are not rare, with the 2+2 hierarchy (two binaries orbiting a common center of mass) being the most frequent configuration\footnote{For a catalog of low-mass stars in multiples, see the Multiple Star Catalog (\url{ http://www.ctio.noao.edu/~atokovin/stars/index.html}). See also tables in \citet{sana2014} and \citet{sana2017}, specifically for massive (O- and B-type) stars in multiple systems.} For instance, \citet{rid15} found a $\sim 5\%$ abundance of 2+2 quadruples. Just like triple systems, quadruples can undergo KL cycles, but they have a larger portion of the phase space where excursions to high eccentricity can occur \citep{pejcha2013,grishlai2018}. As a consequence, even though quadruples are rarer, the fraction of systems that that produce mergers is higher compared to triples \citep{fragk2019,liu2019}.

In a recent paper, \citet{safar2020} proposed that two episodes of KL-induced mergers would first cause two NSs to merge and form a low-mass-gap BH, which
can subsequently merge with another BH in a 3+1 quadruple. However, even a small recoil kick for the first NS--NS merger remnant could possibly unbind the outer orbits, thus preventing a second merger. Moreover, the 3+1 systems they considered are typically less common in nature than the 2+2 hierarchies, by a factor of a few \citep[e.g.,][]{tok14a,tok14b}. In this Letter, we show that 2+2 systems can lead to BH mergers in both the low- and high-mass gap. In our scenario, the BH in the mass gap (resulting from the first merger, of either two NSs or two BHs) is imparted a recoil kick, which triggers its interaction with the second binary in the system (see Figure~\ref{fig:interq}). The outcome of the interaction, as we show below, will often be a new binary containing the BH in the mass gap and merging within a Hubble time with another BH.

The paper is organized as follows. In section~\ref{sect:bhgap} we discuss the formation and recoil of BHs in the low- and high-mass gaps within 2+2 quadruples. In Section~\ref{sect:bhgap} we provide a numerical demonstration of the proposed mechanism, and, in Section~\ref{sect:rate}, we discuss how to estimate the merger rate of such objects. Finally, we discuss the model and draw our conclusions in Section~\ref{sect:conc}.

\section{Black holes in the low- and high-mass gaps in 2+2 quadruples}
\label{sect:bhgap}

We start by describing the basic steps that lead to the production of BHs in the low- and high-mass gaps in 2+2 quadruples.

To produce a 2+2 system of compact objects, each of the two stellar binaries in the progenitor quadruple has to be stable against dynamical perturbations by the companion binary. This can be ensured by requiring the 2+2 system to satisfy the stability criterion for hierarchical triples derived in, e.g., \citet{mar01}, assumed to be valid for quadruple systems if the third companion is appropriately replaced by a binary system,
\begin{equation}
\frac{A_{\rm out}}{A_{\rm in}}\geq \frac{2.8}{1-E_{\rm out}} \left[\left(1+\frac{m_{\rm out}}{m_{\rm in}}\right)\frac{1+E_{\rm out}}{\sqrt{1-E_{\rm out}}} \right]^{2/5}\ .
\label{eqn:stabts}
\end{equation}
Here, $A_{\rm out}$ and $E_{\rm out}$ are the semi-major axis and eccentricity of the progenitor outer orbit, $m_{\rm out}$ the total mass of the progenitor binary companion, and $m_{\rm in}$ and $A_{\rm in}$ the total mass and the orbital semi-major axis of the progenitor binary that we require to be stable.

As discussed in \citet{safar2020}, quadruples can in principle be disrupted by occasional flybys with other stars in the field \citep{ha2018}. This process occurs over an evaporation timescale \citep{binntrem87}
\begin{eqnarray}
T_{\rm EV}&=&3\times 10^2\ \mathrm{Gyr}\left(\frac{v_{\rm disp}}{20\kms}\right)\left(\frac{0.6\msun}{\langle m_*\rangle}\right)\times \nonumber\\
&\times& \left(\frac{0.1\msun\ \mathrm{pc}^{-3}}{\rho}\right)\left(\frac{100\ \mathrm{AU}}{A_{\rm out}}\right)\left(\frac{M_{\rm tot}}{100\msun}\right)\ ,
\end{eqnarray}
where $v_{\rm disp}$ is the stellar velocity dispersion, $\langle m_*\rangle$ the average perturber mass, $\rho$ the stellar density, and $M_{\rm tot}$ the total progenitor quadruple mass. The catastrophic regime where the system is disrupted by a single encounter takes place on longer timescales. From equation~(2) we see that the relevant timescale for the disruption of a typical quadruple considered in the present work is of the order of hundreds of Gyrs and therefore flybys can safely be neglected.

Quadruple systems can experience significant KL oscillations already on the main sequence, which could drive them to merge prematurely during this phase whenever the KL cycles are not damped by relativistic or tidal precession \citep{shapp2013,michp2014,fang2018}. Phases of Roche-lobe overflow and common envelopes can occur in each of the two binaries of the quadruple in the exact same way it happens for isolated binary stars (assuming no interaction between the two widely separated binaries). This can ultimately lead the two massive binary stars to evolve to become two compact-object binaries. The exact evolution could be much more complicated if episodic mass loss occurs due to eccentric Roche-lobe overflow and/or if common envelope phases in the quadruple were to happen on timescales comparable to the KL oscillations \citep{hamd2019,rosa2019}.

\begin{figure} 
\centering
\includegraphics[scale=0.565]{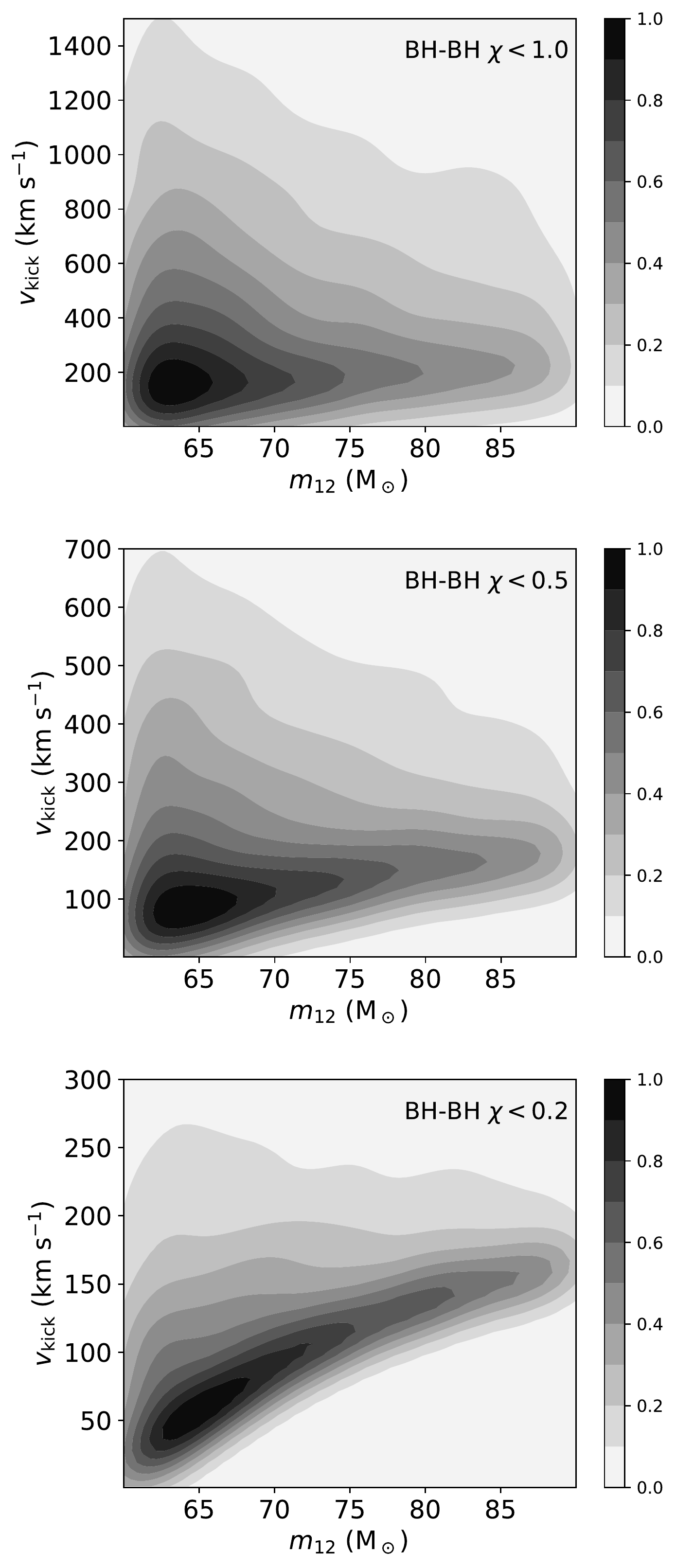}
\caption{Density distributions of the recoil kick velocity imparted to the remnant of the merger of two BHs with masses $\ge 30\msun$, following the results of \citet{lou10}. The masses of the two BHs are drawn from a power law $m^{-2.3}$ and are paired following a uniform mass-ratio distribution. The reduced spins are uniformly distributed with magnitudes $<1$ (top), $<0.5$ (center), $<0.2$ (bottom). The spin directions are assumed uniform on the sphere.}
\label{fig:vkick}
\end{figure}

After main-sequence lifetime is over, massive stars explode to form a compact objects. After every explosive event, the system is imparted a kick as a result of the mass loss \citep{bla1961} and a natal kick due to recoil from an asymmetric supernova explosion. The latter typically follows a Maxwellian distribution, with a characteristic velocity dispersion $\sigma$. The value of $\sigma$ is highly uncertain, and can be $\sim 100\kms$ \citep{arz2002} or as high as $\sim 260\kms$ \citep{hobbs2005} for NSs. On the other hand, the kick can be as low as zero for electron-capture SNe \citep{pod2004}. For BHs, a common assumption is that the momentum imparted to a BH is the same as the momentum given to a NS, assuming momentum conservation \citep{fryer2001}. As a consequence, the kick velocities for BHs should be typically lower by a factor of $m{_{\rm NS}}/ m{_{\rm BH}}$ with respect to NSs ($m{_{\rm NS}}$ and $m{_{\rm BH}}$ are the NS and BH mass, respectively).

\begin{table}
\caption{Model parameters: name, mass of the remnant from the merger of $m_1$ and $m_2$ ($m_{12}$), primary mass in the companion binary ($m_3$), secondary mass in the companion binary ($m_4$), semi-major axis of the companion binary ($a_{34}$), recoil kick velocity ($\vkick$).}
\centering
\begin{tabular}{lccccc}
\hline
Name & $m_{12}$ ($\msun$) & $m_3$ ($\msun$) & $m_4$ ($\msun$) & $a_{34}$ (AU) & $\vkick$ ($\kms$)\\
\hline\hline
Low1 & 3 & 30 & 1.4 & 1  & 10\\
Low2 & 3 & 50 & 1.4 & 1  & 10\\
Low3 & 3 & 30 & 5   & 1  & 10\\
Low4 & 3 & 30 & 10  & 1  & 10\\
Low5 & 3 & 30 & 1.4 & 10 & 10\\
Low6 & 3 & 30 & 1.4 & 1  & 50\\
\hline\hline
High1 & 70  & 30 & 30 & 1  & 50 \\
High2 & 100 & 30 & 30 & 1  & 50 \\
High3 & 70  & 50 & 50 & 1  & 50 \\
High4 & 70  & 50 & 30 & 1  & 50 \\
High5 & 70  & 30 & 30 & 10 & 50 \\
High6 & 70  & 30 & 30 & 1  & 100\\
\hline
\end{tabular}
\label{tab:models}
\end{table}

To model self-consistently the stellar evolution of a population of 2+2 quadruple systems is not straightforward since the two binaries are not isolated and their evolutionary pathways could be quite unusual. For example, eccentric mass transfer or a common envelope that enshrouds the whole quadruple system could occur. Tools to handle these situations have not been developed yet. However, we can estimate the number of 2+2 systems that can form a quadruple of compact objects and lead to a NS--NS or BH--BH merger. Based on \citet{sana2012} we assume that each progenitor binary in the quadruple follows a distribution of periods
\begin{equation}
\mathcal{F}(P)\propto \log_{10}^{-0.55} (P/1\,{\rm d}),
\end{equation}
and eccentricities
\begin{equation}
\mathcal{F}(e)\propto e^{0.42}.
\end{equation}
We use the results of \citet{gm2018}, who showed that a majority of binaries that produce NS--NS and BH--BH remnants have final distribution of semimajor axes peaking around $10$--$10^3\rsun$, depending on the common-envelope parameters and the natal kicks: with larger $\sigma$ the peak of the semimajor axis distribution shifts to lower values. The outer orbit has to be stable against the systemic velocity imparted to binaries as a result of SN kicks. To ensure this, the natal kicks have to satisfy \citep{kalog1996}
\begin{equation}
v_{\rm sys}\lesssim 40\ {\kms} \left(\frac{\mu_{\rm Q}}{40\msun}\right)^{1/2} \left(\frac{A_{\rm out}}{100\ {\rm AU}}\right)^{-1/2}\ ,
\label{eqn:vkicknatal}
\end{equation}
where $\mu_{\rm Q}$ is the progenitor quadruple's reduced mass. We find that $\sim 0.01\%$--$0.1\%$, $\sim 0.1\%$--$1\%$, $\sim 1\%$--$10\%$ of the quadruples survive natal kicks with $\sigma=260\kms$, $100\kms$, $20\kms$, respectively.

After a quadruple of compact objects is formed (see Figure~\ref{fig:interq}) and is stable according to Eq.~\ref{eqn:stabts}, BHs and NSs in each of the two binaries in the 2+2 system can merge either because the common envelope phase left them with small enough separations \citep{bel16b} or as a result of the KL mechanism \citep{fragk2019,liu2019}. \citet{fragk2019} showed that, even though quadruples are rarer, the fraction of systems that merge is higher with respect to triples owing to a more complex dynamics. The merger remnant is imparted a recoil kick owing to asymmetries at the moment of the merger. The recoil kick for BH--BH mergers depends on the mass ratio and the spins of the merging objects. For NS--NS mergers, its magnitude could be much smaller since the encounter takes place at a larger gravitational radius, but hydrodynamic effects could become important \citep{lou10,lou11,rezz2010}. As an example, we show in Figure~\ref{fig:vkick} the density distributions of the recoil kick velocity $v_{\rm kick}$ imparted to the remnant of the merger of two BHs with masses $\ge 30\msun$, following \citet{lou10}. The masses of the two BHs are drawn from a simple power law $\propto m^\gamma$, with $\gamma=-2.3$, and are paired following a uniform mass-ratio distribution. Note that \citet{fish2020} have shown that current LIGO/Virgo BH detections are consistent with $\gamma\approx -1.1$ and that the two BHs within each merging binary tend to have comparable masses. However, these early results may be affected by selection effects and may change with the upcoming results from LIGO/Virgo O3 run. The dimensionless spin parameter\footnote{Defined as $a_K/m < 1$, where $a_K$ is the usual Kerr parameter.} is assumed uniformly distributed with magnitude $<1$ (top), $<0.5$ (center), or $<0.2$ (bottom). The spin directions are assumed to be isotropic. While for high spins $v_{\rm kick}$ can be as high as about $1400\kms$, systems that merge with low spins have a maximum recoil kick around $250 \kms$, with the bulk near $50\kms$.

\begin{figure*} 
\centering
\includegraphics[scale=0.575]{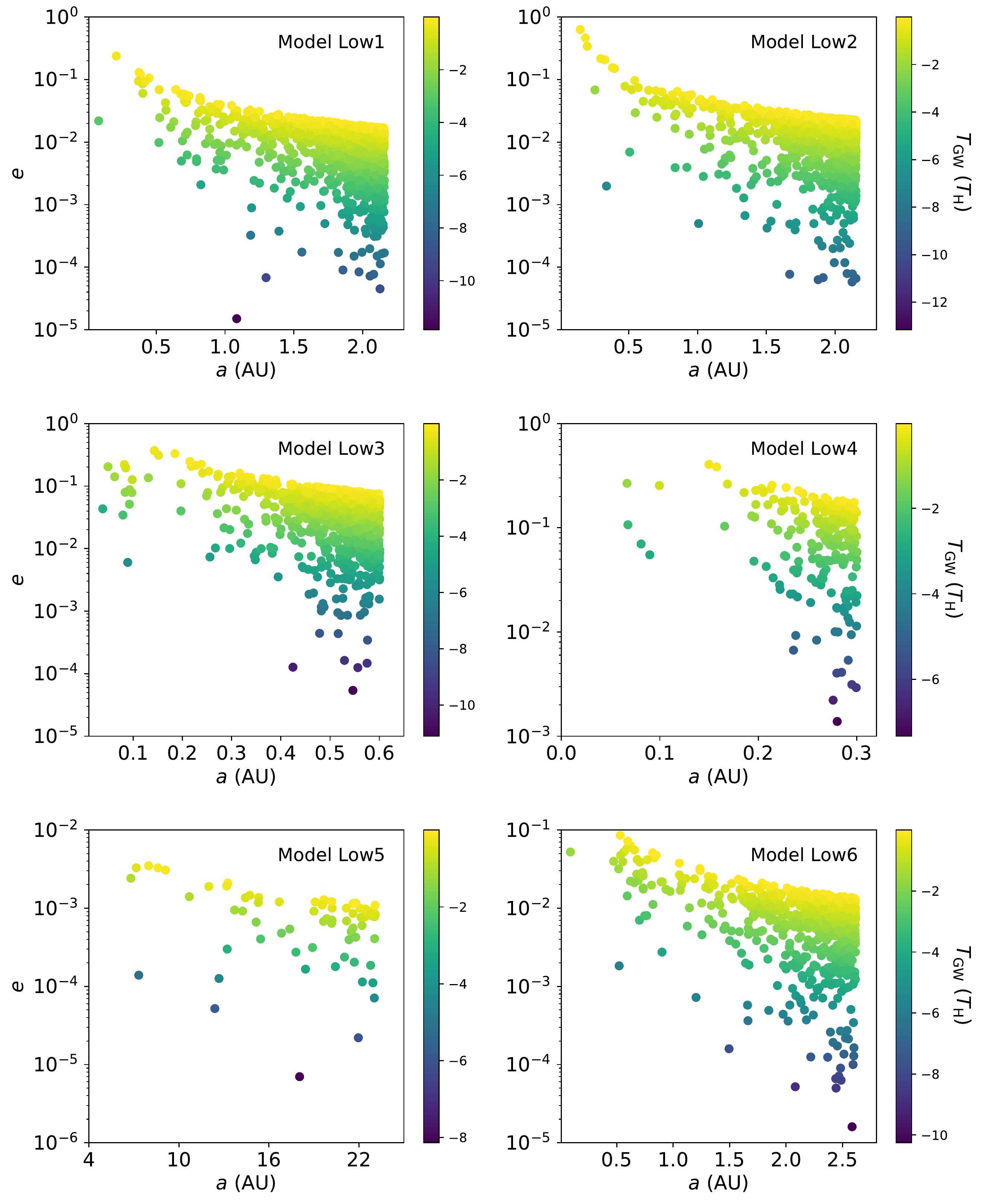}
\caption{Distribution of semi-major axes ($a$) and eccentricities ($e$) of the binary systems that contain a low-mass-gap BH that merge in a Hubble time, for the 6 different models in Table~\ref{tab:models}. These merging systems are formed through the mechanism discussed in Section~\ref{sect:bhgap}.}
\label{fig:lowmass}
\end{figure*}

\begin{figure*} 
\centering
\includegraphics[scale=0.575]{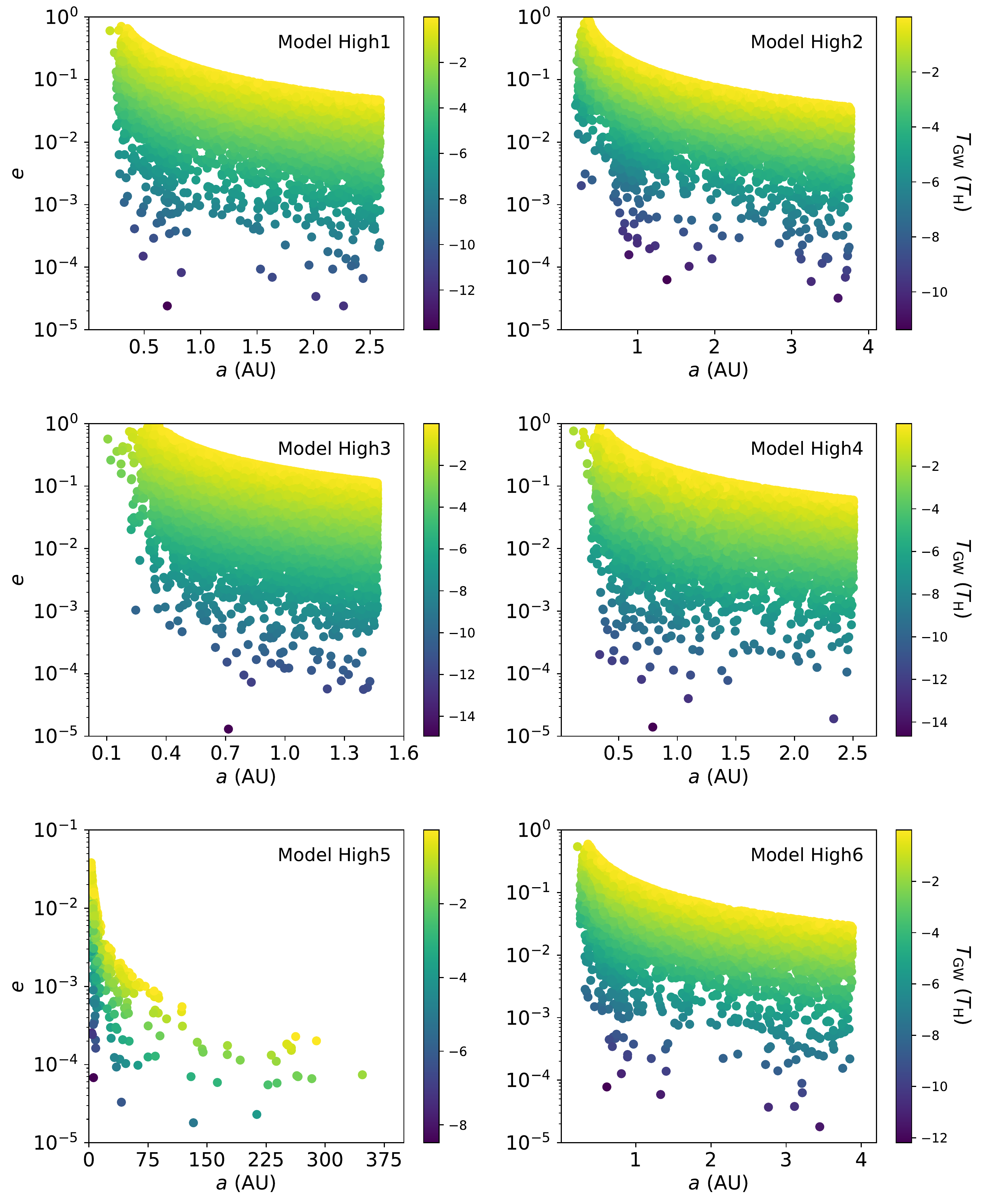}
\caption{Distribution of semi-major axes ($a$) and eccentricities ($e$) of the binary systems that contain a high-mass-gap BH merging within a Hubble time, for the 6 different models in Table~\ref{tab:models}.}
\label{fig:highmass}
\end{figure*}

To avoid the system recoiling into a stable triple \citep[e.g. see][]{fraloeb2019}, the recoil kick velocity has to be roughly larger than the outer orbital speed of the 2+2 system,
\begin{equation}
v_{\rm kick}\gtrsim 10\ {\kms} \left(\frac{\mu}{10\msun}\right)^{1/2} \left(\frac{a_{\rm out}}{100\ {\rm AU}}\right)^{-1/2}\ ,
\label{eqn:vkicklow}
\end{equation}
where $\mu=m_{12}m_{34}/m_{\rm tot}$ is the quadruple reduced mass, and $m_{12}=m_1+m_2$ and $m_{34}=m_3+m_4$ (see Figure~\ref{fig:interq}). To ensure a resonant encounter and that the outcome of the encounter of $m_{12}$ against the binary companion ($m_3$-$m_4$) is a binary system, the kick velocity should not be much larger than the $m_3$-$m_4$ orbital speed,
\begin{equation}
v_{\rm kick}\lesssim 100\ {\kms} \left(\frac{\mu_{34}}{10\msun}\right)^{1/2} \left(\frac{a_{34}}{1\ {\rm AU}}\right)^{-1/2}\ ,
\label{eqn:vkickhigh}
\end{equation}
where $\mu_{34}=m_{3}m_{4}/(m_3+m_4)$ is the reduced mass of the companion binary. With these conditions satisfied, the velocity kick vector also has to lie in the fractional solid angle,
\begin{equation}
\mathcal{F}\sim \left(\frac{B a_{\rm in}}{a_{\rm out}}\right)^2\ ,
\label{eqn:solidangle}
\end{equation}
where $B>1$ is the gravitational focusing factor in the scattering cross-section,
\begin{equation}
\sigma \sim \pi B^2 a_{34}^2 \sim \pi a_{34}^2 \left(\frac{2Gm_{\rm tot}}{a_{34}v_{\rm kick}^2} \right)\ ,
\label{eqn:focus}
\end{equation}
where $m_{\rm tot}$ is the total quadruple mass.

\section{Numerical example}
\label{sect:example}

In this Section, we provide a numerical example of the above scenario producing a merger in the low- or high-mass-gap from a 2+2 quadruple system. For simplicity, we consider the system after the formation of all compact objects\footnote{It would be sufficient that either $m_3$ or $m_4$ is a compact object. In this case, collisions with non-compact stars may occur in resonant encounters.}, thus ignoring the details of the quadruple evolution before the formation of BHs and NSs. Many effects (natal kicks, common envelope phases, mass transfer from winds or Roche-lobe overflow, etc.) could be significant and some fraction of the quadruple population will not survive. We leave detailed calculations of all these effects to future work, and simply demonstrate that a BH merger in the low- and high-mass gaps is possible in 2+2 quadruples, whenever their stellar progenitors can successfully produce a quadruple of compact objects. 

Further, we assume that the merger of two NSs (BHs) has produced a BH in the low- (high-) mass gap of mass $m_{12}$ (see Fig.~\ref{fig:interq}), which is imparted a recoil kick $v_{\rm kick}$ and interacts with the components of the other binary, with component masses $m_3$ and $m_4$. We use the \textsc{FEWBODY} numerical toolkit for computing these 1+2 close encounters \citep{fregeau2004}, which can result in a new binary containing a BH in either the low- or high-mass gap, that could later merge within a Hubble time. We take into account the different masses of the compact objects involved in the interaction, the semi-major axis $a_{34}$ of the binary (assumed to be on a circular orbit), and different recoil kick velocities (see Eqs.~\ref{eqn:vkicklow}-\ref{eqn:vkickhigh}). The impact parameter is drawn from a distribution
\begin{equation}
f(b)=\frac{b}{2b_{\rm max}^2}\ ,
\end{equation}
where $b_{\rm max}$ is the maximum impact parameter of the scattering experiment defined in Eq.~\ref{eqn:focus}. We study 12 different models, 6 for the merger of a BH in the low-mass gap and 6 for the merger of a BH in the high-mass gap (Table~\ref{tab:models}). We run $10^5$ integrations for each model, for a total of $1.2\times 10^6$ integrations.

We show in Figure~\ref{fig:lowmass} the distribution of semi-major axes ($a$) and eccentricities ($e$) of the binary systems that contain a low-mass-gap BH ($3\msun$) merging within a Hubble time, for the 6 different models (Low1-6) in Tab.~\ref{tab:models}. In these runs, we fix $m_3$ as the BH primary, while $m_4<m_3$ is taken to be as a NS or a secondary BH. We find that both resonant and non-resonant encounters can produce binaries containing a BH in the low-mass gap. In Model Low1 ($m_3=30\msun$, $m_4=1.4\msun$, $a_{34}=1$ AU, $v_{\rm kick}=10\kms$), the fraction of binaries that merge after formation is $1.4\times 10^{-2}$. These systems have typical initial semi-major axis $\lesssim 2.2$ AU and eccentricity $\lesssim 0.1$. We find that a larger $m_3$ mass (Model Low2) does not significantly affect the properties and the fraction of merging binaries, while they change for more massive secondary $m_4$ masses (Models Low3-4). In this case, merging binaries are formed with smaller semi-major axes and the merging fraction is $2.3\times 10^{-3}$ for $m_4=10\msun$. Larger values of $a_{34}$ and $v_{\rm kick}$ decrease the fraction of merging systems to $7.3\times 10^{-4}$ and $7.1\times 10^{-3}$, respectively, with the former also producing wider merging binaries.

In Figure~\ref{fig:highmass}, we show the distribution of semi-major axes ($a$) and eccentricities ($e$) of the binary systems that contain a high-mass-gap BH merging in a Hubble time, for the 6~different models (High1-6) in Table~\ref{tab:models}. We consider $m_3$ and $m_4$ as the BH primary and secondary, respectively. Also in this case, we find that both resonant and non-resonant encounters can produce binaries containing a BH in the high-mass gap. In Model High1 ($m_{12}=70\msun$, $m_3=m_4=30\msun$, $a_{34}=1$ AU, $v_{\rm kick}=10\kms$), the fraction of binaries that merge after formation is $1.2\times 10^{-1}$. Typical initial semi-major axes are $\lesssim 2.7$ AU and the binaries can even be formed circular unlike the case of the low-mass-gap mergers, owing to the larger masses involved in the scenario. A larger $m_{12}$ (Model High2) does not affect the properties and the merger fraction of the binaries being formed, while larger $m_3$ and $m_4$ masses (Models High3-4) produce more compact binaries and the fraction of mergers increases to $1.7\times 10^{-1} - 2.4\times 10^{-1}$. As in the low-mass-gap case, larger values of $a_{34}$ and $v_{\rm kick}$ decrease the fraction of merging systems to $4.1\times 10^{-3}$ and $9.3\times 10^{-2}$, respectively.

\section{Merger rate}
\label{sect:rate}

In this Letter, our goal is to present a new possible pathway to form merging BHs in the low- and high-mass gap. The difficulty in modeling self-consistently the stellar evolution in a population of 2+2 quadruple systems comes mainly from having two binaries that are not isolated and can be strongly affected by KL oscillations \citep{shapp2013,michp2014,fang2018}. For example, phases of Roche-lobe overflow and common envelopes can occur with eccentric orbits (on timescales comparable to the KL oscillations), unlike the typical case for isolated binaries \citep{hamd2019,rosa2019}. Even more complicated would be episodes of mass transfer between the two binaries, or a whole-quadruple common-envelope phase. Nevertheless, we can derive an order of magnitude estimate for the merger rates of BHs in the low- and high-mass gap from our proposed scenario.

The LIGO detector horizon for NS--NS mergers is $\sim 120(M_{\rm chirp}/1.2 \msun)^{5/6}$ Mpc, where $M_{\rm chirp}$ is the chirp mass of the system. Assuming that a BH in the mass gap merges with a BH of mass $\sim 30\msun$, a merger event in the low-mass (high-mass) gap has a detection horizon of $\sim 1$ Gpc ($40$ Gpc). Any mechanism producing a merger rate $\gtrsim 0.01$ Gpc$^{-3}$ yr$^{-1}$ could lead to  detections within the next decade \citep{safar2020}.

We adopt an average star formation rate of $10^8\msun$ Gpc$^{-3}$ yr$^{-1}$ \citep{madau2014}. A number of authors have shown that BH--BH and NS--NS mergers have an efficiency of one merger per $\sim 10^{-5}$--$10^{-6} \msun$ depending on natal kicks, common-envelope efficiency, and metallicity \citep[e.g.,][]{bel16b,gm2018}. In the case of a quadruple, this efficiency can be increased because of KL cycles. This leads to a merger rate of $\sim 10^2$--$10^3$ Gpc$^{-3}$ yr$^{-1}$. We now account for the fact that $\sim 0.01\%$--$0.1\%$, $\sim 0.1\%$--$1\%$, $\sim 1\%$--$10\%$ of the quadruples survive natal kicks with $\sigma=260\kms$, $100\kms$, $20\kms$, respectively. Large natal kicks are expected for NSs, except when they were born from an electron-capture process, while low natal kicks are expected for BHs, particularly of high mass, as a result of momentum conservation and fallback. Considering that $\sim 5\%$ of massive stars are in 2+2 systems, the merger rate in quadruples can be estimated as $\sim 10^{-4}$--$10$ Gpc$^{-3}$ yr$^{-1}$. This simple estimate is roughly consistent with the results of \citet{fragk2019}, who showed that the merger rate from quadruples could be comparable to that from triple systems \citep{ra2018,fk2020} owing to a more complex dynamics, even though quadruples are rarer.

For the parameters we have explored in our numerical experiments (see Table~\ref{tab:models}), we have found that the fraction of systems that merge in the low- and high-mass gap is $\sim 10^{-3} - 10^{-2}$ and $\sim 10^{-3} - 10^{-1}$, respectively. Therefore, the merger rate from our proposed mechanism would be $\sim 10^{-7} - 10^{-2}$ Gpc$^{-3}$ yr$^{-1}$ and $\sim 10^{-3} - 10^{-2}$ Gpc$^{-3}$ yr$^{-1}$ for BHs in the low-mass and high-mass gap, respectively. This estimate could likely be affected by accounting for the proper kick velocity magnitude (Eqs.~\ref{eqn:vkicklow}-\ref{eqn:vkickhigh}) and the relevant processes (as Roche-lobe overflow, common envelope phases, etc.) that shape the evolutionary paths of 2+2 quadruples systems. We leave detailed calculations of their possible effect to future work.

\section{Conclusions}
\label{sect:conc}

In this Letter, we have shown that 2+2 quadruple systems can lead to BH mergers in the low- ($\lesssim 5\msun$) and high-mass gap ($\gtrsim 50\msun$). In our scenario, the BH in the mass gap originates from the merger of two NSs or BHs in one of the two binaries of the quadruple and is imparted a kick velocity, which triggers its interaction with the second binary of the system. The outcome of the encounter will eventually be a new binary containing the BH in the mass gap and merging with a new BH companion within a Hubble time. We have demonstrated how this mechanism works by considering different masses of the compact objects involved in the interaction, different semi-major axes of the companion binary in the 2+2 quadruple, and different recoil kick velocities. We have shown that smaller recoil kicks and larger primary masses in the companion binary produce a larger number of merging BHs in the low- and high-mass gaps. We have also estimated that the merger rate from our proposed mechanism is $\sim 10^{-7} - 10^{-2}$ Gpc$^{-3}$ yr$^{-1}$ and $\sim 10^{-3} - 10^{-2}$ Gpc$^{-3}$ yr$^{-1}$ for BHs in the low-mass and high-mass gap, respectively.

Interestingly, our proposed mechanism could also account for much more extreme mass ratios in the merging BH--BH binaries, unlike those in mergers from isolated binaries or cluster dynamics, which are never far from unity \citep{bel16b,rod18}. We also predict that these systems (BH mergers in the mass gaps) would appear very nearly circular in the LIGO-Virgo frequency band, contrary to the KL-induced mergers in hierarchical triple systems \citep[e.g.][]{fk2020}. Finally, this scenario can also produce electromagnetic counterparts whenever the components of the second binary (which interact with the first merger product) are not both BHs. In that case a merging BH--NS binary can be produced, or a non-compact star could collide with one of the BHs during the interaction \citep{fraglpk2019,kremer2019}.

Our proposed scenario, while promising, is certainly not unique. Moreover, mergers with objects in the low-mass gap may simply indicate a delayed core-collapse engine \citep{fryer2012}, while mergers of objects in the high-mass gap may simply reflect our limited understanding of stellar evolution \citep{woosley2017,limongi2018,bel2020}. As the detector sensitivity is improved, hundreds of merging binary signals are expected to be detected by LIGO-Virgo in the next few years, and tighter constraints will be placed on the BH mass function, thus shedding light on the low- and high-mass gaps, and testing in some detail the various formation mechanisms for populating them \citep{kove2017,fish2019}.

\section*{Acknowledgements}

GF thanks Tsvi Piran, Philipp Moesta, and Kyle Kremer for useful discussions, and the anonymous referee for a constructive report. GF acknowledges support from a CIERA Fellowship at Northwestern University. AL was supported in part by Harvard's Black Hole initiative, which is funded by JTF and GBMF grants. FAR acknowledges support from NSF Grant AST-1716762.

\bibliographystyle{yahapj}
\bibliography{refs}

\end{document}